\documentclass[twocolumn,aps,showpacs,amsmath,amssymb,floatfix,superscriptaddress]{revtex4}
\usepackage{graphicx}% Include figure files 
\usepackage{dcolumn}% Align table columns on decimal point 
\usepackage{bm}% bold math
\usepackage{hyperref} 
\usepackage{latexsym}
\usepackage{float}

\begin{document}
\title{Ultimate Fate of Constrained Voters} 
\author{F. Vazquez}\email{fvazquez@buphy.bu.edu}
\author{S.~Redner}\email{redner@bu.edu}
\affiliation{Center for BioDynamics, Center for Polymer Studies, 
and Department of Physics, Boston University, Boston, MA, 02215}

\begin{abstract}
  
  We determine the ultimate fate of individual opinions in a
  socially-interacting population of leftists, centrists, and rightists.  In
  an elemental interaction between agents, a centrist and a leftist can
  become both centrists or both become leftists with equal rates (and
  similarly for a centrist and a rightist).  However leftists and rightists
  do not interact.  This interaction step between pairs of agents is applied
  repeatedly until the system can no longer evolve.  In the mean-field limit,
  we determine the exact probability that the system reaches consensus
  (either leftist, rightist, or centrist) or a frozen mixture of leftists and
  rightists as a function of the initial composition of the population.  We
  also determine the mean time until the final state is reached.  Some
  implications of our results for the ultimate fate in a limit of the Axelrod
  model are discussed.

\end{abstract}  


\maketitle

\section{Introduction}
  
A basic issue in social dynamics is to understand how opinion diversity
arises when interactions between individuals are primarily ``ferromagnetic''
in character.  Many kinetic spin models have been proposed to address this
general question \cite{G,LN97,W,Sz,KR}.  An important example of this genre
is the appealingly simple Axelrod model \cite{axelrod, CMV}, which accounts
for the formation of distinct cultural domains within a population.  In the
Axelrod model, each individual is endowed with a set of features (such as
political leaning, music preference, choice of newspaper, {\it etc.}), with a
fixed number of choices for each feature.  Evolution occurs by the following
voter-model-like update step \cite{voter}.  A random individual is picked and
this person selects an interaction partner (anybody in the mean-field limit,
and a nearest-neighbor for finite spatial dimension).  For this pair of
agents a feature is randomly selected.  If these agents have the same state
for this feature, then another feature is picked and the initial person
adopts the state of this new feature from the interaction partner.  This
dynamics mimics the feature that individuals who share similar sentiments on
lifestyle issues can have a meaningful interaction in which one will convince
the other of a preference on an issue where disagreement exists.

Depending on the number of traits and the number of states per trait, a
population may evolve to global consensus or it may break up into distinct
cultural domains, in which individuals in different domains do not have any
common traits \cite{axelrod, CMV}.  This diversity is perhaps the most
striking feature of the Axelrod model.  An even simpler example with a
related phenomenology is the bounded compromise model \cite{Deff,HK,BKR}.
Here, each individual possesses a single real-valued opinion that evolves by
compromise.  In an update step, two interacting individuals average their
opinions if their opinion difference is within a pre-set threshold.  However,
if this difference is greater than the threshold, there is no interaction.
These steps are repeated until the system reaches a final state.  For a
sufficiently large threshold the final state is consensus, while for a
smaller threshold the system breaks up into distinct opinion clusters; these
are analogous to the cultural domains of the Axelrod model.

In spite of the simplicity of these models, most of our understanding stems
from simulation results.  As a first step toward analytic insight, a discrete
three-state version of the bounded compromise model was recently introduced
\cite{VKR}.  This is perhaps the simplest opinion dynamics model that
includes the competing features of consensus and incompatibility.  This model
was found to exhibit a variety of anomalous features in low dimensions,
including slow non-universal kinetics and power-law spatial organization of
single-opinion domains.  In this work, we focus on the ultimate fate of this
system in the mean-field limit.  We determine the exact probability that the
final state of the system is either consensus or a frozen mixture of leftists
and rightists as a function of the initial composition of the system.  We
also compute the time required for the system to reach its ultimate state.

In the next section, we define our model, outline its basic properties, and
determine the ultimate fate of the system in terms of an equivalent
first-passage process.  The solution to this problem is used to obtain the
probabilities of reaching either a frozen final state (Sec.~III) or consensus
(Sec.~IV).  In Sec.~V, we compute the mean time until the final state is
reached.  In the discussion section, we generalize to non-symmetric
interactions and also show how our results can be adapted to determine the
ultimate fate of a version of the Axelrod model.  Calculational details are
presented in the appendices.

\section{The Model}

\begin{figure}[ht] 
 \vspace*{0.cm}\hspace*{0.1in}
 \includegraphics*[width=0.35\textwidth]{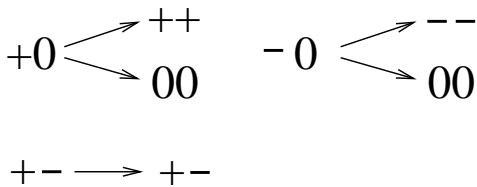}
\caption{Update events for different pair states.}
\label{steps}
\end{figure}

We consider a population of $N$ individuals, or agents, that are located at
the nodes of a graph.  Each agent can be in one of three opinion states:
leftist, centrist, or rightist. As shown in Fig.~\ref{steps}, we represent
these states as $-, 0$, and $+$, respectively.  In a single microscopic event
an agent is selected at random.  We consider the mean-field limit in which
the neighbor of an agent can be anyone else in the system.  If the two agents
have the same opinion, nothing happens.  If one is a centrist and the other
is an extremist (or vice versa), the initial agent adopts the opinion of its
neighbor --- that is, each individual can be viewed as having zero
self-confidence and merely adopts the state of a compatible neighbor; this
kinetic step is the same as in the classical voter model \cite{voter}.
However, if the two agents are extremists of opposite persuasions, they are
incompatible and do not influence each other.  As a result of this
incompatibility constraint, the final state of the system can be either
consensus of any of the three species, or a frozen mixture of leftist and
rightist extremists, with no centrists.

In a mean-field system with $N_-$ leftists , $N_+$ rightists, and $N_0$
centrists (with $N_+ + N_- + N_0 =N$), the probability of selecting a pair
$ij$ ($i,j=+,-,0$) equals $2N_i N_j/[N(N-1)]$, where $N_i$ is the number of
agents of type $i$.  Then the elemental update steps and their respective
probabilities are:
\begin{eqnarray}
\label{hops}
(N_-,N_0)\to(N_-\!\pm\! 1,N_0\!\mp\! 1) &{\rm prob.}&
p_x=\frac{N_-N_0}{N(N-1)}\nonumber \\
(N_+,N_0)\to(N_+\!\pm\! 1,N_0\!\mp\! 1) &{\rm prob.}& 
p_y=\frac{N_+N_0}{N(N-1)},\nonumber\\
\end{eqnarray}
while the probability for no change is $1-2(p_x+p_y)$.

\begin{figure}[ht] 
 \vspace*{0.cm}
 \includegraphics*[width=0.40\textwidth]{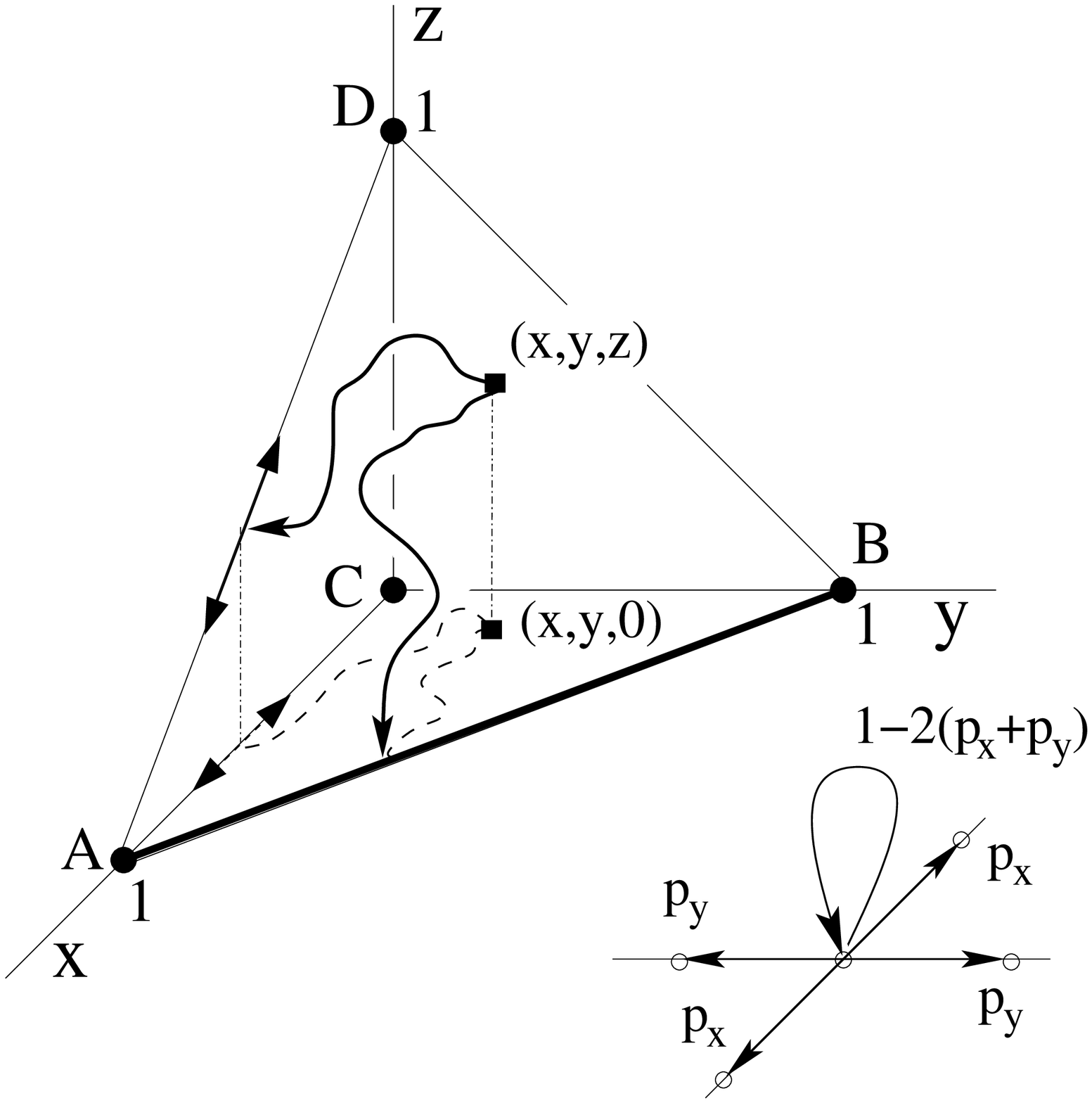}
\caption{The density triangle $x+y+z=1$.  
  The heavy dots denote consensus states and the heavy line denotes the
  frozen final states where no centrists remain.  Typical random walk
  trajectories are shown, along with their projections (dashed) on the $z=0$
  plane.  When a random walk reaches the lines $x=0$ or $y=0$, the random
  walk subsequently must remain on this line until consensus is reached.  The
  corresponding hopping probabilities on the $z=0$ plane are also indicated.}
\label{3d-triangle}
\end{figure}

Eventually the system reaches a static final state that is either consensus
of one of the three species or a frozen mixture of leftists and rightists.
Monte Carlo simulations showed that the nature of the final state has a
non-trivial dependence on the initial densities of the species \cite{VKR}.
We now analytically determine this final state probability as a function of
the initial population composition by solving an equivalent first-passage
problem.  As shown in Fig.~\ref{3d-triangle}, the state of the system
corresponds to a point in the space of densities $x=N_-/N$, $y=N_+/N$, and
$z=N_0/N$.  Since $x+y+z=1$, this constraint restricts the system to the
triangle $ABD$ shown in the figure.

When two agents interact, the state of the system may change and we can view
this change as a step of a corresponding random walk on the triangle $ABD$,
with single step hopping probabilities given in Eq.~(\ref{hops}).  When the
walk reaches one of the fixed points $A$, $B$, or $D$ (consensus), or any
point on the fixed line $AB$ (frozen extremist mixture), the system stops
evolving.  This set represents absorbing boundaries for the effective random
walk.  Thus to find the probability of reaching a given final state, we
compute the first-passage probability for the random walk to hit a given
absorbing boundary.

We may simplify the problem by using the fact that only two of the densities
$(x,y,z)$ are independent.  We thus choose $x$ and $y$ as the independent
variables by projecting the effective random walk trajectory onto the $z=0$
plane (Fig.~\ref{3d-triangle}).  According to Eq.~(\ref{hops}), this
two-dimensional random walk jumps to its nearest neighbors in the $x$ and $y$
directions with respective probabilities $p_x=xz/(1-\delta)\approx xz$ and
$p_y=yz/(1-\delta)\approx yz$, and stays in the same site with probability
$1-2(p_x+p_y)$.  Here $\delta=1/N$, and we consider the limit $N\to\infty$.

\section{Frozen Final State}

We now determine the probability $P_{+-}(x,y)$ for the system to reach a
frozen state when the initial densities are $\rho_-(t=0)\equiv x$ and
$\rho_+(t=0)\equiv y$.  This coincides with the first-passage probability for
the equivalent random walk to hit the line $x+y=1$ when it starts at a
general point $(x,y)$ in the interior of the triangle.  There are also
absorbing boundaries on the sides $x=0$ and $y=0$, where the probability of
reaching the frozen state is zero, and $P_{+-}(x,y)=1$ on the line segment
$x+y=1$.  For notational simplicity, we write $F(x,y)\equiv P_{+-}(x,y)$.  In
the equivalent random walk process, this first-passage probability obeys the
recursion \cite{redner}
\begin{eqnarray}
\label{Fxy}
F(x,y)&=& p_x[F(x-\delta,y)+F(x+\delta,y)]\nonumber \\
 &+& p_y[F(x,y-\delta)+F(x,y+\delta)] \nonumber \\
 &+& [1-2(p_x+p_y)]F(x,y). 
\end{eqnarray}
That is, the first-passage probability $F(x,y)$ equals the probability of
taking a step in some direction (the factor $p_i$) times the first-passage
probability from this target site to the boundary.  This product is then
summed over all possible target sites after one step of the walk.

In the continuum limit ($N \rightarrow \infty$) we expand this equation to
second order in $\delta$ and obtain
\begin{equation}
 x\frac{\partial^2 F(x,y)}{\partial x^2}+y\frac{\partial^2 F(x,y)}
 {\partial y^2}=0,
 \label{FP_eq}
\end{equation}
supplemented with the boundary conditions
\begin{eqnarray*}
&& F(x,0)=0 \nonumber \\
&& F(0,y)=0 \nonumber \\
&& F(x,1-x)=1.
 \label{BC0}
\end{eqnarray*}
The solution to Eq.~(\ref{FP_eq}) is not straightforward because of the mixed
boundary conditions.  Instead of attacking the problem directly, we transform
to the coordinates $u=\sqrt x$, $v=\sqrt y$ to map the triangle to a
quarter-circle of unit radius and then apply separation of variables in this
geometry.

The transformed differential equation is
\begin{eqnarray}
 \frac{\partial^2 F(u,v)}{\partial u^2}&+&\frac{\partial^2 F(u,v)}
{\partial v^2} \nonumber \\
&-&\frac{1}{u}\frac{\partial F(u,v)}{\partial{u}}-
 \frac{1}{v}\frac{\partial F(u,v)}{\partial{v}}=0.
 \label{uv_eq}
\end{eqnarray}
Because of the circular symmetry of the boundary conditions in this reference
frame, it is now convenient to use the polar coordinates $(\rho,\theta)$,
where $u=\rho \cos \theta$ and $v=\rho \sin \theta$.  This transforms
Eq.~(\ref{uv_eq}) to
\begin{eqnarray}
 \frac{\partial^2 F(\rho,\theta)}{\partial \rho^2}&-&
 \frac{1}{\rho}\frac{\partial F(\rho,\theta)}{\partial{\rho}}
 +\frac{1}{\rho^2}\frac{\partial^2 F(\rho,\theta)}{\partial{\theta^2}}
 \nonumber \\
&+& \frac{1}{\rho^2}
 \left(\tan\theta-\cot\theta\right)
 \frac{\partial F(\rho,\theta)}{\partial \theta}=0
 \label{PC}
\end{eqnarray}
We seek a product solution $F(\rho,\theta)=R(\rho)\Theta(\theta)$.
Substituting this into Eq.~(\ref{PC}) leads to the separated equations
\begin{eqnarray}
 \label{R}
&& \frac{d^2 R}{d \rho^2}-\frac{1}{\rho}\frac{d R}{d\rho}-
 \frac{m^2}{\rho^2}R=0 \\
  \nonumber \\
 \label{Theta}
&& \frac{d^2 \Theta}{d\theta^2}+\left(\tan\theta-
 \cot\theta\right) \frac{d \Theta}{d\theta}+m^2\Theta=0
\end{eqnarray}
where $m$ is the separation constant. 

Eq.~(\ref{R}) is equidimensional and thus has the general power-law form
\begin{eqnarray}
\label{R-gen}
 R(\rho)=A_+ \rho^{1+\sqrt{1+m^2}}+A_- \rho^{1-\sqrt{1+m^2}}
\end{eqnarray}
where $A_\pm$ are constants.  To solve Eq.~(\ref{Theta}), we proceed by
eliminating the first derivative to give a Schr\"{o}dinger-like equation.
Thus we define $\Theta(\theta)\equiv f(\theta)\,u(\theta)$ and find the
function $f(\theta)$ that eliminates this first derivative term.

Substituting $\Theta=f\,u$ in Eq.~(\ref{Theta}) gives
\begin{eqnarray}
 fu''&+&\left[2f'+\left(\tan\theta-
 \cot\theta \right)f\right]u' \nonumber \\
 &+&\left[f''+\left(\tan\theta-\cot\theta\right)f'+m^2f \right]u=0,
 \label{Sch0}
\end{eqnarray}
and the coefficient of $u'$ is zero if $f$ satisfies
\begin{eqnarray}
 2 f'+ \left(\tan\theta-\cot\theta\right) f=0, \nonumber
\end{eqnarray}
whose solution is
\begin{equation}
\label{f}
f(\theta)\propto \sqrt{2\sin\theta\cos\theta} \propto \sqrt{\sin 2\theta}.
\end{equation}
Substituting this expression for $f$ in Eq.~(\ref{Sch0}), we obtain
\begin{equation}
 \frac{d^2u}{d\theta^2}+\left[(1+m^2)
 -\frac{3}{4}\left(\frac{1}{\sin^2\theta}+\frac{1}{\cos^2\theta}\right) \right]u=0.
 \label{Sch1}
\end{equation}

Details of the solution to this Schr\"odinger equation are given in Appendix
A.  The final result is
\begin{eqnarray}
\label{F-polar}
F(\rho,\theta)&=&R(\rho)f(\theta)u(\theta)\nonumber \\
&=&\sum_{n~ {\rm odd}} \frac{(2n+1)}{n(n+1)} \; \rho^{2(n+1)}\,
 \sin 2 \theta \,\,
 P_{n}^1 \left(\cos 2\theta\right),\nonumber\\
\end{eqnarray}
where $P_n^1$ is the associated Legendre function.  We transform back to the
original $xy$ coordinates through
\begin{eqnarray*}
 u= \rho \cos \theta=\sqrt{x} \qquad
 v= \rho \sin \theta=\sqrt{y} \\
 \sin 2 \theta=\frac{2 \sqrt{x y}}{x+y} \qquad
 \cos 2 \theta=\frac{x-y}{x+y}
\end{eqnarray*}
Finally, identifying $F$ with $P_{+-}$, the solution in the original
Cartesian coordinates is
\begin{equation}
 P_{+-}(x,y)=\sum_{n~ {\rm odd}} \frac{2(2n+1)}{n(n+1)} \, \sqrt{xy}\, (x+y)^n \, 
 P_n^1\left(\frac{x-y}{x+y}\right).
 \label{sol1}
\end{equation}

This gives the probability of reaching a final frozen state when the
population has initial densities $x$ and $y$ of leftists and rightists,
respectively.  Notice that this probability is symmetric,
$P_{+-}(x,y)=P_{+-}(y,x)$, because $P_n^1$ is an even function for $n$ odd
and an odd function for $n$ even.  This reflects the obvious physical
symmetry that the probability $P_{+-}$ is invariant under the interchange of
leftists and rightists.

\begin{figure}[ht] 
 \vspace*{0.cm}
\vskip 0.05in
 \includegraphics*[width=0.42\textwidth]{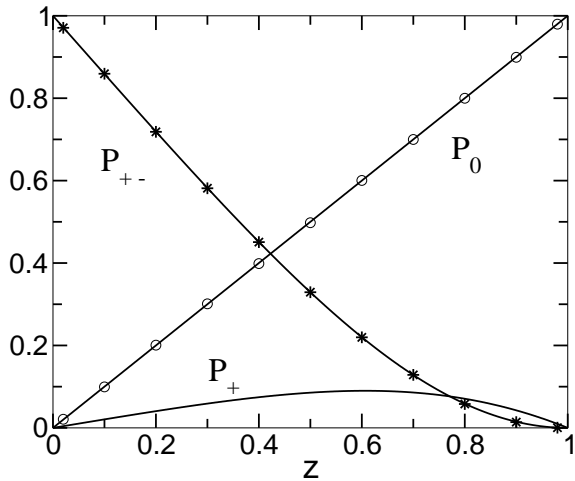}\vskip 0.2in
 \includegraphics*[width=0.42\textwidth]{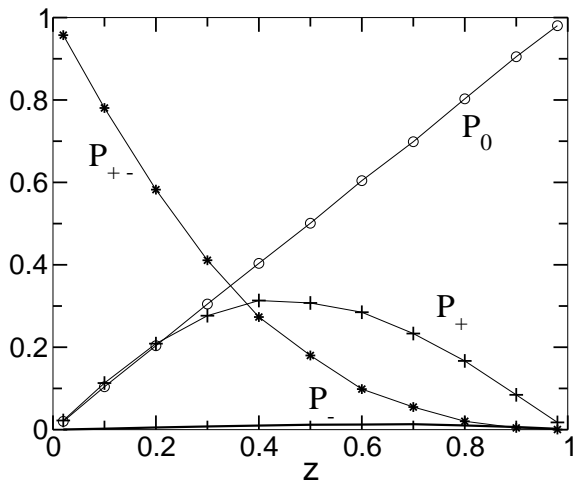}
\caption{Probability of reaching the frozen final state, $P_{+-}$, and
  extremist consensus, $P_+$, as a function of the initial density of
  centrists $z$ for the case of equal initial densities of leftists and
  rightists, $y/x=1$ (top) and for the case $y/x=9$ (bottom).}
\label{solution}
\end{figure}

For equal initial densities of leftists and rightists, namely $x=y=(1-z)/2$,
Eq.~(\ref{sol1}) becomes a function of the initial density of centrists $z$
only
\begin{eqnarray*}
\label{P-closed}
 P_{+-}(z)=\sum_{n~ {\rm odd}} \frac{(2n+1)}{n(n+1)} (1-z)^{n+1} 
 \; P_n^1(0).
\end{eqnarray*}
As shown in Appendix B, this can be simplified to the closed-form expression
\begin{equation}
 P_{+-}(z)=1-\frac{1-(1-z)^2}{\sqrt{1+(1-z)^2}}
 \label{final-sol}
\end{equation}
with $0 \leq z \leq 1$.  This solution is shown in Fig.~\ref{solution} along
with Monte-Carlo simulation results.

\begin{figure}[ht]
 \vspace*{0.cm}
 \includegraphics*[width=0.4\textwidth]{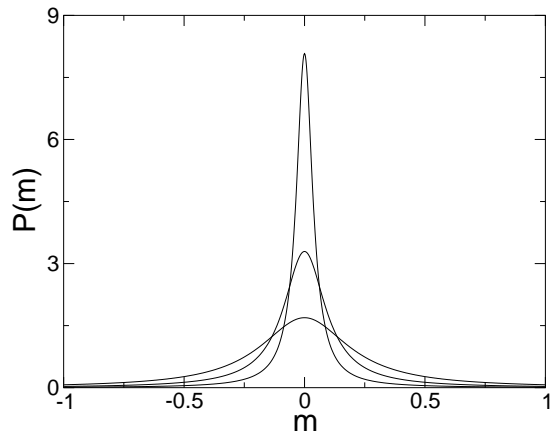}
 \caption{Probability distribution that an initially symmetric 
   system $(x=y)$ has magnetization $m$ in the frozen final state, for the
   cases $z=0.04$, 0.1, and 0.2 (progressively broadening curves).}
 \label{magnet}
\end{figure}

\section{Consensus Final States}

In addition to reaching a frozen final state, the system can also reach
global consensus --- either leftist, rightist, or centrist.  Again, the
probabilities for these latter three events depends on the initial
composition of the system.  We thus define $P_-(x,y)$, $P_+(x,y)$, and
$P_0(x,y)$, as the respective probabilities to reach a leftist, rightist, or
centrist consensus as a function of the initial densities.

The probability for centrist consensus can be obtained by elementary means.
We merely map the 3-state constrained system onto the 2-state voter model by
considering the leftist and rightist opinions as comprising a single
extremist state, while the centrist opinion state maintains its identity.
The dynamics of this system is exactly that of the classical 2-state voter
model.  Since the overall magnetization of the voter model is conserved
\cite{voter}, the probability that the system reaches centrist consensus, for
a given $x$ and $y$ is simply
\begin{equation}
\label{P0}
P_0(x,y)=z=1-(x+y).
\end{equation}

To find the probability for leftist and rightist consensus, we use
normalization of the final state probability
\begin{eqnarray}
\label{norm}
 P_+(x,y)+P_-(x,y)+P_{+-}(x,y)+P_0(x,y)=1
\end{eqnarray}
and conservation of the global magnetization to write
\begin{eqnarray}
\label{mag-norm}
y-x&=& P_+(x,y)-P_-(x,y) \nonumber \\   &+&\int_0^1 (1-2x')F(x'|x,y) dx'.
\end{eqnarray}
Here $F(x'|x,y)$ is the probability of ending in a specific frozen state with
density $x'$ of $-$ spins and $1-x'$ of $+$ spins as a function of the
initial densities $x$ and $y$.  This function therefore satisfies the
normalization condition $P_{+-}(x,y)=\int_0^1 F(x'|x,y)dx'$.  Additionally
the integral in Eq.~(\ref{mag-norm}) is the final magnetization in the frozen
state.  From the exact solution for $F(x'|x,y)$ given in Eq.~(\ref{Fx'}), we
thereby obtain as a byproduct the magnetization distribution in the final
frozen state (Fig.~\ref{magnet}).  As expected, for a small initial density
of centrists $z$, there is little evolution before the final state is reached
and the magnetization distribution is narrow.

Using Eq.~(\ref{P0}) and the normalization condition for $F(x'|x,y)$, we
recast Eqs.~(\ref{norm}) and (\ref{mag-norm}) as
\begin{equation}
 P_+(x,y)+P_-(x,y)+P_{+-}(x,y)=x+y,
 \label{Prob-cons}
\end{equation}
and
\begin{eqnarray}
 P_+(x,y)&-&P_-(x,y)+P_{+-}(x,y)\nonumber \\
&-&2\int_0^1x'F(x'|x,y) dx'=y-x.
 \label{Mag-cons}
\end{eqnarray}
Subtracting these equations, we obtain
\begin{equation}
 P_-(x,y)=x-\int_0^1 x'F(x'|x,y)dx'
 \label{P+}
\end{equation}

Now the first-passage probability $F(x'|x,y)$ to a frozen state with
a specified density $x'$ of $-$ spins obeys the same differential equation as
$P_{+-}$ [Eq.~(\ref{FP_eq})], but with the boundary conditions
\begin{eqnarray}
\label{BC-Fx'}
 F(x'|x,0)&=&0 \nonumber \\
 F(x'|0,y)&=&0 \nonumber \\
 F(x'|x,1-x)&=&\delta(x-x')
 \label{BC2}
\end{eqnarray}
The last condition states that the first-passage probability to the point
$(x',1-x')$ on the boundary $x+y=1$ is zero unless the effective random walk
starts at $(x',1-x')$.  The first-passage probability $F(x'|x,y)$ has the
general form given in Eq.~(\ref{sol0}), but with the coefficients $c_n$ now
determined from the boundary conditions in Eq.~(\ref{BC2}).

From the solution for $F(x'|x,y)$ given in Appendix A, the probability of $-$
consensus as a function of the initial densities $x$ and $y$ is
\begin{equation}
 P_-(x,y)
 \label{P+1}
 =x-\sum_{n=1}^\infty \frac{(2n+1)}{n(n+1)} \, \sqrt{xy}\, (x+y)^n \,  
 P_n^1\left(\frac{x-y}{x+y}\right)
\end{equation}
The probability of $+$ consensus can be obtained by using the fact that
$P_+(x,y)=P_-(y,x)$.

For the special case of $x=y=(1-z)/2$, the probabilities
$P_+(z)=P_-(z)$ can be obtained either by setting $x=y$ in
Eq.~(\ref{P+1}) and summing the series, or, more simply, by using
Eq.~(\ref{final-sol}) for $P_{+-}(z)$ and the probability conservation
equation $ 2P_+(z)+P_{+-}(z)+z=1$.  By either approach, we
find the following closed expression for the probability of extremist
consensus as a function of $z$
\begin{equation}
 P_+(z)=P_-(z)=\frac{1}{2}\left(\frac{1-(1-z)^2}
 {\sqrt{1+(1-z)^2}}-z \right).
 \label{P+2}
\end{equation}
This result is also shown in Fig.~\ref{solution}.

\begin{figure}[ht]
 \vspace*{0.cm}
 \includegraphics*[width=0.35\textwidth]{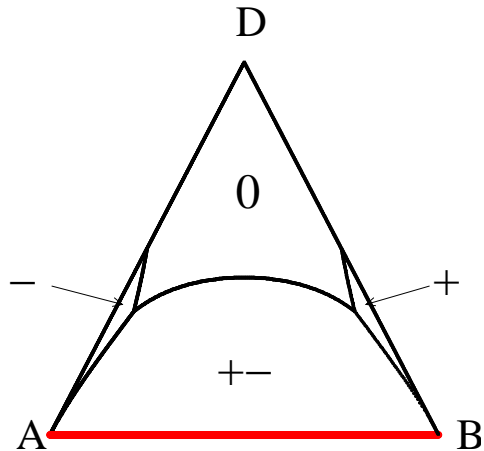}
 \caption{Exact phase diagram in the composition triangle $ABD$ 
   of Fig.~\ref{3d-triangle}.  The triple points, where the final state
   probabilities of three of the four phases are equal, are located at
   $(x,y)\approx(0.0507,0.6185)$ and $(x,y)\approx(0.6185,0.0507)$.}
 \label{phase-diag}
\end{figure}

From our results for $P_{+-}(x,y)$, $P_+(x,y)$, $P_-(x,y)$, and $P_0(x,y)$,
we may define a ``phase diagram'' shown in figure~\ref{phase-diag}.  Each
region in the figure corresponds to the portion of the composition triangle
where the probability of ultimately ending up in the labeled state is
greater than all other first-passage probabilities.  The least likely outcome
is the achievement of extremist consensus while the most likely result is to
get stuck in the frozen mixed state.

\section{MEAN EXIT TIME}

In addition to the probability of reaching a particular final state, we also
study the mean time until the final state is reached as a function of the
initial composition of the system.  The simplest such quantity is the {\em
  unconditional\/} mean time $t(x,y)$ to reach {\em any} of the four possible
final states --- extremist consensus ($+$ or $-$), centrist consensus, and
mixed frozen state, as a function of the initial densities $x$ and $y$.  This
first-passage time can again be obtained trivially by considering an
equivalent two-state system where we lump leftists and rightists into a
single extremist state.  The system stops evolving when the density of
centrists reaches $z=0$ or $z=1$, with the former corresponding either to
extremist consensus or to a frozen mixed state.

To find this first-passage time to reach the final state, note that in a
single event the effective one-dimensional random walk that corresponds to
the state of the system can either jump to one of its two nearest neighbors
with probability $p_z$ or stay at the same site with probability $1-2p_z$,
where $p_z=p_x+p_y=z(x+y)=z(1-z)$.  The time interval for each event is
$dt=1/N$, corresponding to each person being selected once on average every
$N$ update steps.  Then the mean time to reach the final states $z=0$ or
$z=1$ as a function of the initial density $z$ obeys the recursion
\cite{redner}
\begin{eqnarray}
 t(z)&=&p_z[t(z+\delta )+dt]+p_z[t(z-\delta )+dt] \\
     &+&(1-2p_z)[t(z)+dt].
\end{eqnarray}
This formula has a similar form and a similar explanation as the equation for
the first-passage probability [Eq.~(\ref{Fxy})].  Starting from $z$, the mean
time to reach the final state equals the probability of taking a single step
(the factors $p_z$ and $1-2p_z$) multiplied by the time needed to reach the
boundaries via this intermediate site.  This path-specific time is just the
mean first-passage time from the intermediate site plus the time $dt$
for the initial step.

In the large-$N$ limit this recursion reduces to
\begin{eqnarray}
\label{tz}
 \frac{d^2t(z)}{d z^2}=-\frac{1}{D},
\end{eqnarray}
where the diffusion coefficient is $D={p_z}\delta^2/dt$.  Since $dt=\delta
=1/N$, we have $D= z(1-z)/N$.  Eq.~(\ref{tz}) is subject to the boundary
conditions $t(0)=t(1)=0$, corresponding to immediate absorption if the random
walk starts at the boundary.  The solution to this equation is
\begin{equation}
 t(z)=-N \left[ z \; \ln z + (1-z) \; \ln(1-z) \right].
 \label{time}
\end{equation} 
This is simply the mean consensus time of the 2-state voter model in the
mean-field limit, in which the initial density of the two species are $z$ and
$1-z$.

\begin{figure}[ht]
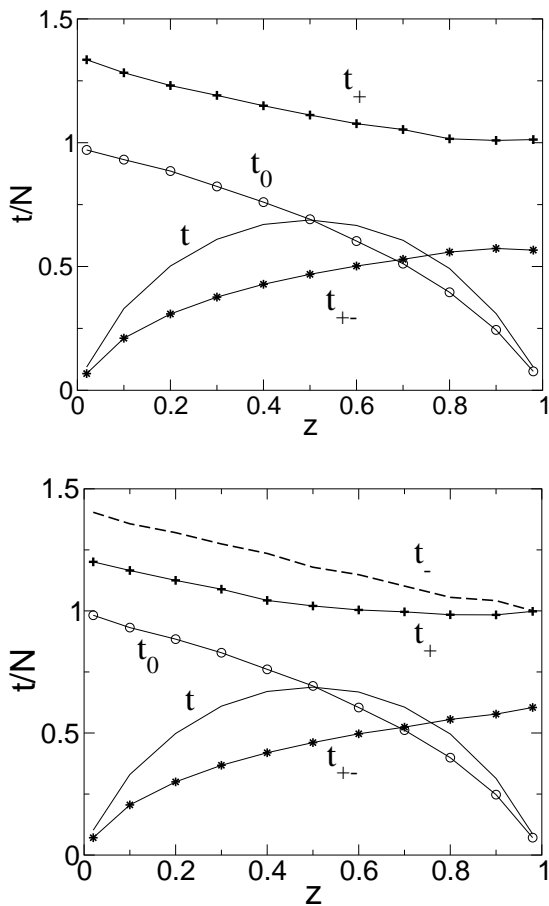

 \vspace*{0.cm}
 \includegraphics*[width=0.4\textwidth]{time_r_1.eps}\vskip 0.2in
 \includegraphics*[width=0.4\textwidth]{time_r_2.eps}
 \caption{Unconditional and conditional exit times as a function of $z$ for
   the cases $y/x=1$ (top) and $y/x=2$ (bottom).  The curves for $t$ and $t_0$
   are based on exact formulae, while the other are based on simulations.
   The latter results are based on $10^5$ realizations for each initial
   state. }
 \label{times}
\end{figure}

It is more interesting to consider the {\em conditional\/} first-passage time
to a specific final state as a function of the initial condition.  For
example, consider the conditional time $t_0(z)$, the mean time to reach
centrist consensus when the initial centrist density is $z$.  This is the
mean time for the equivalent random walk to hit the point $z=1$, {\em
  without\/} hitting $z=0$.  In the large $N$ limit, this conditional exit
time obeys the differential equation (see Sec.~1.6 in \cite{redner})
\begin{eqnarray}
\label{t-cond}
D \frac{d^2}{dz^2}[P_0(z)\;t_0(z)]= -P_0(z).  
 \label{t+} 
\end{eqnarray} 
Integrating Eq.~(\ref{t-cond}), using $P_0(z)=z$, and the boundary conditions
$P_0(z)\;t_0(z)=0$ at $z=0$ and at $z=1$, we obtain $P_0(z)\;t_0(z)=-N \;
(1-z) \; \ln (1-z)$, from which the conditional time to reach centrist
consensus is
\begin{eqnarray}
 t_0(z)=-N\,\frac{1-z}{z} \; \ln (1-z).
\end{eqnarray} 
Similarly, the mean time to reach all other final states of the system is
\begin{equation}
 t_{\rm other}(z)=-N\,\frac{z}{1-z} \; \ln z.
\end{equation}

In principle, we can find the conditional times to reach the extremist
consensus states and the frozen final state as a function of the initial
condition.  This involves solving the two-dimensional analogue of
Eq.~(\ref{t-cond}) in the density triangle, subject to the appropriate
boundary conditions.  Because of the tedious nature of the calculation, we
have instead resorted to numerical simulations to compute these conditional
exit times.  These results are shown in Fig.~\ref{times}.  While all exit
times are of the order of $N$, it is worth noting that $t_+$ and $t_-$ are
typically the longest times.  This stems from the fact that reaching
extremist consensus is a two-stage process.  First, the extremists of the
opposite persuasion must be eliminated, and then there is a subsequent
first-passage process in which the centrists are also eliminated.

\section{Discussion}

We determined basic properties of the final state in a simple opinion
dynamics model that consists of a population of leftists, rightists, and
centrists.  Centrists interact freely with extremists (either leftist or
rightist), so that one agent adopts the opinion of its interaction partner,
while extremists of opposite persuasions do not interact.  These competing
tendencies lead either to ultimate consensus or to a frozen final state that
consists of non-interacting leftists and rightists.  While this model is
clearly an oversimplification of opinion evolution in a real population, it
provides a minimalist description for how consensus or distinct cultural
domains can be achieved.

Our calculations are based on the mean-field limit in which each individual
interacts with any other individual with equal probability.  It is worth
emphasizing that the dependence of the final state probabilities on the
initial composition of the population is very close to the corresponding
quantities on finite-dimensional systems \cite{VKR}.  This coincidence is a
reflection of the conservation of the global magnetization by the dynamics.
Indeed, for the classical voter model, the conservation of the magnetization
immediately leads to the fact that the probability of $+$ consensus is equal
to the initial density of $+$ spins in any dimension.

There are two extensions of the model that are worth mentioning.  First, it
is natural that extremists have a stronger conviction than centrists.  This
suggests the generalization where a centrist adopts an extremist's opinion
with probability $p>1/2$ while an extremist adopts a centrist's opinion with
probability $q=1-p$ in an elementary update step.  For this generalization,
the densities of each species evolve by the following rate equations in the
mean-field limit
\begin{eqnarray*}
 \label{1}
 \dot x&=& v \,z\, x \\ 
 \label{2}
 \dot y&=& v \,z \,y \\
 \label{3} 
 \dot z&=& -v \,z\,(x+y) 
\end{eqnarray*}
where $v=p-q$ quantifies the bias.  For $p\ne q$, The solution to these
equations are
\begin{eqnarray*}
 x(t)=\frac{x(0) \; e^{v t}}{1+[x(0)+y(0)](e^{v t}-1)},
\end{eqnarray*}
and similarly for $y(t)$, while $z(t)=1-x(t)-y(t)$.  In the long time limit
and for $p>q$, we find
\begin{eqnarray*}
 x  \rightarrow  \frac{x(0)}{x(0)+y(0)},\qquad
 y  \rightarrow  \frac{y(0)}{x(0)+y(0)},
\end{eqnarray*}
while $z \rightarrow 0$.  Thus if there is an innate bias favoring
extremism, a frozen final state with no consensus is inevitable.  On the
other hand, if centrists are dominant ($p<q$), there centrist consensus is
the final result.

Our results can also be adapted to determine the final states probabilities 
in the mean-field limit of the Axelrod model for the case of two features and 
two states per feature. The four possible states of an individual can be 
represented by

\begin{eqnarray*}
++\qquad --\qquad +-\qquad -+\;.
\end{eqnarray*}
We define the dynamics to be that if an interaction pair has one
common feature, then the other feature where disagreement exists is 
picked and the initial agent takes on the state of the interaction partner.

There are four possible consensus states: all $--$, $-+$, $+-$, and $++$, and
two types of frozen states: mixtures of $--$ and $++$, and mixtures of $-+$
and $+-$.  We can identify the states $--$ and $++$ in this Axelrod model
with the $-$ and $+$ states, respectively, in our spin-1 Ising model, and the
states $-+$ and $+-$ taken together with the 0 state of the spin-1 Ising
model. Then, using our previous results, if the initial densities of $--$ and
$++$ states are $x$ and $y$ respectively, the probabilities of $--$ and $++$
consensus and the probability of frozen mixture of $--$ and $++$ are given by
the expressions $P_-(x,y)$, $P_+(x,y)$ and $P_{+-}(x,y)$ respectively (Eqs.
(\ref{P+1}) and (\ref{sol1})).  By similar arguments, the probabilities of
$-+$ and $+-$ consensus and the probability of $-+$ and $+-$ frozen mixture
are given by the above expressions where $x$ and $y$ are now the initial
densities of $-+$ and $+-$ states respectively.  Unfortunately, this exact
mapping onto an equivalent Ising spin system does not work when there are
more than two features and/or more than two states per feature.

\section{Acknowledgments}

We thank Paul Krapivsky for useful discussions and helpful advice.  We also
thank NSF grant DMR0227670 for financial support of this research.

\newpage
\appendix
\onecolumngrid
\section{Solution to Equivalent Schr\"odinger Equation}

We present the solution to Eq.~(\ref{Sch1}).  This turns out be the
Schr\"{o}dinger equation for the \emph{P\"oschl-Teller potential hole}
\cite{QM-book}, for which the generic form of the equation is
\begin{equation}
 \frac{d^2u(\theta)}{d\theta^2}+
 \left[k^2-\alpha^2 \left(\frac{\chi(\chi-1)}{\sin^2\alpha \theta}+
 \frac{\lambda(1-\lambda)}{\cos^2\lambda \theta}\right)\right]u(\theta)=0,
 \label{Sch}
\end{equation}
with $k^2=1+m^2$, $\alpha=1$, $\chi=\lambda$, and $\chi(\chi-1)=3/4$.  This
last equation has the roots $\chi_1=3/2$ and $\chi_2=-1/2$.  The eigenvalues
and eigenvectors of Eq.~(\ref{Sch}) are (see, for example, \cite{QM-book}
page 92):
\begin{eqnarray*}
 k^2 &=&\alpha^2(\chi+\lambda+2 n)^2 \\
 u_n(\theta)&=& \sin^\chi \alpha \theta \, \cos^\lambda \alpha \theta \;
 _2F_1(-n,\chi+\lambda+n, \chi+\frac{1}{2}; \sin^2\alpha \theta), 
\end{eqnarray*}
with $n=0,1,2,..$, and where $_2F_1$ is the hypergeometric function.

For $\chi=-1/2$, $u_n(\theta)$ diverges at $\theta=0$ and $\theta=\pi/2$.
Thus we take the solution for $\chi=3/2$ only and obtain the eigenfunctions
and eigenvectors of Eq.~(\ref{Sch1}):
\begin{eqnarray}
 \label{eme}
 1+m^2 &=&(3+2 n)^2 \\
 u_n(\theta)&=& \sin^\frac{3}{2} \theta \, \cos^\frac{3}{2} \theta \,\,\,
 _2F_1(-n,3+n, 2; \sin^2 \theta). \nonumber
\end{eqnarray}\\
The solutions for the angular function in Eq. (\ref{Theta}) therefore have
the form
\begin{eqnarray*}
 \Theta_n(\theta)= f(\theta)\, u_n(\theta)=\sin^2 2\theta\,\,
 _2F_1(-n,3+n, 2; \sin^2 \theta),
\end{eqnarray*}
with $m$ given by Eq. (\ref{eme}).  Finally, the generic form of the product
solution $F_n(\rho,\theta)=R_n(\rho)\, \Theta_n(\theta)$ to Eq. (\ref{PC}) is
\begin{equation}
 F_n(\rho,\theta)=\left(A_+ \rho^{2(2+n)}+A_- \rho^{-2(1+n)}\right)
 \Theta_n(\theta). \nonumber
\end{equation}
The coefficient $A_-$ must be zero because $F$ is zero at the origin.  Thus
\begin{equation}
 F_n(\rho,\theta)=\rho^{2(2+n)} \sin^2(2 \theta) \,\,
 _2F_1(-n,3+n,2;\sin^2\theta), \nonumber
\end{equation} 
and the general solution to Eq. (\ref{PC}) is
\begin{equation}
 F(\rho,\theta)=\sum_{n=0}^\infty c_n \; \rho^{2(2+n)}\; \sin^2 2\theta \,\,
 _2F_1(-n,3+n,2;\sin^2\theta).
 \label{sol}
\end{equation} 

We need the set of hypergeometric functions $_2F_1$ to form an orthogonal set
to obtain the coefficients $c_n$ from the boundary condition:
\begin{equation}
 F(\rho=1,\theta)=1=\sum_{n=0}^\infty c_n \;\sin^2 2\theta \,\,
 _2F_1(-n,3+n,2;\sin^2\theta). \nonumber
\end{equation}
Fortunately, the functions $_2F_1$ are related, for certain specific integer
parameters, to the Associated Legendre Polynomials (that are known to form an
orthogonal set) via \cite{Grad}:
\begin{eqnarray*}
 P_\nu^m(x)=\frac{(-1)^m \;\Gamma(\nu+m+1)\; (1-x^2)^{m/2}}{2^m \;
 \Gamma(\nu-m+1)\;m!}\;
 _2F_1(m-\nu,m+\nu+1;m+1;\frac{1-x}{2}).
\end{eqnarray*}
If we take $m=1$, $\nu=n+1$ and $x=\cos 2\theta$ we obtain:
\begin{eqnarray*}
 _2F_1(-n,n+3;2;\sin^2\theta)=\frac{-2 n!}{(n+2)!\; \sin 2\theta}\;
 P_{n+1}^1(\cos 2\theta).
\end{eqnarray*}  
Then the general solution Eq. (\ref{sol}) can be written as
\begin{equation}
 F(\rho,\theta)=\sum_{n=0}^\infty c_n \; \rho^{2(2+n)}\; \sin 2\theta \,\,
 P_{n+1}^1(\cos 2\theta),
 \label{sol0}
\end{equation}
where we absorbed the factor $\frac{-2 n!}{(n+2)!}$ into the coefficient
$c_n$.

\subsection{Solution for $P_{+-}$}

To determine $P_{+-}$, we now apply the boundary condition $F=1$ at $\rho=1$
and obtain
\begin{eqnarray*}
\label{F-prelim}
 F(\rho=1,\theta)=1=\sum_{n=0}^\infty c_n \; \sin 2\theta \,
 P_{n+1}^1(\cos 2\theta),
\end{eqnarray*} 
and using the orthogonality-normalization of the $P_n^1$'s we obtain 
the $c_n$'s:
\begin{eqnarray}
\label{ON}
c_n = \left\{ \begin{array}{ll}
       {\displaystyle  0} & \mbox{~~~~~ $n$ odd}, \\
      {\displaystyle  \frac{(2n+3)}{(n+2)(n+1)}} & \mbox{~~~~~ $n$ even}, \end{array} \right. 
\end{eqnarray}
Substituting these coefficients in Eq.~(\ref{sol0}), we obtain Eq.~(\ref{F-polar}).

\subsection{Solution for $P_+$}

To determine $P_+$, we first need the first-passage probability $F(x'|x,y)$
for the effective random walker to hit the specific point $(x',1-x')$ on the
locus $x+y=1$, when starting from $(x,y)$.  This first-passage probability
also satisfies the basic differential equation (\ref{FP_eq}), but with the
boundary conditions given in Eq.~(\ref{BC-Fx'}).  In polar coordinates the
last of these boundary conditions is $F(x'|x,1-x)=F(\theta'|\rho=1,\theta)=
\delta(\cos^2\theta-\cos^2\theta')$.  Thus we now determine the coefficients
$c_n$ in Eq.~(\ref{sol0}) by
\begin{eqnarray*}
 \delta(\cos^2\theta-\cos^2\theta')=\sum_{n=0}^\infty c_n \; \sin(2\theta) 
 \,\, P_{n+1}^1(\cos 2\theta).
\end{eqnarray*}
Inverting this relation and using the orthogonality of the $P_n^1$'s
[Eq.~(\ref{ON})], we obtain
\begin{eqnarray*}
 c_n=\frac{(2n+3)}{(n+2)(n+1)} \frac{P_{n+1}^1(\cos 2\theta')}{\sin 2\theta'}.
\end{eqnarray*}
Using this result and transforming back to $xy$ coordinates, we obtain the
first-passage probability
\begin{eqnarray}
\label{Fx'}
 F(x'|x,y)= \sum_{n=1}^\infty \frac{(2n+1)}{n(n+1)} \, 
 \frac{P_n^1(2x'-1)}{\sqrt{x'(1-x')}} \times \sqrt{xy}\, (x+y)^n \,  
 P_n^1\left(\frac{x-y}{x+y}\right).
\end{eqnarray}

We then substitute this expression for $F(x'|x,y)$ into Eq.~(\ref{P+}) to
determine $P_+(x,y)$.  In so doing, we encounter the integral
\begin{eqnarray*}
 I=\int_0^1 x'\frac{P_n^1(2x'-1)}{\sqrt{x'(1-x')}}dx'.
\end{eqnarray*}
To evaluate this integral, we first make the variable change $x=2x'-1$, use
the identity $P_n^1(x)=\sqrt{1-x^2}\; \frac{d P_n(x)}{dx}$, and integrate by
parts to give
\begin{eqnarray}  
\label{ident}
 I&=&\int_{-1}^1 \frac{(x+1)}{2}\,\frac{P_n^1(x)}{\sqrt{1-x^2}}\,dx
= \int_{-1}^1 \frac{(x+1)}{2}\,\frac{dP_n(x)}{dx}\, dx, \nonumber \\
\nonumber \\
 &=& \left. \frac{(x+1)}{2}P_n(x) \right|_{-1}^1-1/2 \int_{-1}^1 P_n(x)\, dx,
 \nonumber\\ \nonumber \\
 &=& P_n(1)-\frac{1}{2} \int_{-1}^1 P_0(x) P_n(x)\, dx=1-\delta_{n0}, 
\end{eqnarray} 
where we have used $P_0(x)=1$, $P_n(1)=1 \;\; \forall n$, as well as 
\begin{eqnarray*}
 \int_{-1}^1 P_m(x)\, P_n(x) \,dx=\frac{2}{2n+1}\,\delta_{mn}.
\end{eqnarray*}
Using the results of Eq.~(\ref{Fx'}) and Eq.~(\ref{ident}) in Eq.~(\ref{P+}),
we obtain the solution quoted in (\ref{P+1}).

\section{Reduction of $P_{+-}$}

To reduce the series representation (\ref{sol1}) for $P_{+-}$ to a
closed form, we start with the identity
\[ P_n^1(0) = \left\{ \begin{array}{ll}
         {\displaystyle 0} & \mbox{$n$ even}, \\
        {\displaystyle \frac{(-1)^{\frac{n-1}{2}} n!!}{(n-1)!!}} & \mbox{$n$ odd}, \end{array} 
        \right. \] 
and make the variable change $n=2m+1$ with $m \geq 0$ integer to obtain
\begin{equation}
 P_{+-}(z)=\sum_{m=0}^\infty \frac{(-1)^m \, (4m+3) \;
 (2m-1)!!}{2(m+1)(2m)!!} \; (1-z)^{2(m+1)}.
 \label{sol_ro}
\end{equation}

This infinite series can be summed by the following algebraic manipulation.
We define $w=(1-z)^2$. The factor $4m+3$ can be rewritten as
$(2m+1)+2(m+1)$, so that Eq.~(\ref{sol_ro}) becomes
\begin{eqnarray*}
 P_{+-}(z) & =& \sum_{m=0}^\infty \frac{(-1)^m \, (2m+1) \; 
 (2m-1)!!}{2(m+1)(2m)!!} \; w^{(m+1)}
            +  \sum_{m=0}^\infty \frac{(-1)^m \, 2(m+1) \;
 (2m-1)!!}{2(m+1)(2m)!!} \; w^{(m+1)}, \\
          &  = & - \sum_{m=1}^\infty \frac{(2m-1)!!}{(2m)!!} \; (-w)^m 
            +  w \sum_{m=0}^\infty \frac{(2m-1)!!}{(2m)!!} \; (-w)^m,\\
           & = & 1-(1-w) \sum_{m=0}^\infty \frac{(2m-1)!!}{(2m)!!}\;(-w)^m, \\
           & = & 1-\frac{1-w}{\sqrt{1+w}},
\end{eqnarray*}
where we used the Taylor series expansion
\begin{eqnarray*}  
  \frac{1}{\sqrt{1-x}} = \sum_{m=0}^\infty \frac{(2m-1)!!}{(2m)!!}\; x^m.
\end{eqnarray*}
This then gives closed expression for the probability of reaching the mixed 
state as a function of $z$ quoted in Eq.~(\ref{P-closed}).

\end{document}